\documentclass[12pt]{article}
\usepackage{graphicx}
\begin{document}

\centerline{\bf Are Barab\'asi-Albert networks self-averaging?}

\centerline{Dietrich Stauffer and Amnon Aharony*}

\centerline{Institute for Theoretical Physics, Cologne
University\\D-50923 K\"oln, Euroland}

* Visiting from
School of Physics and Astronomy, Raymond and Beverly Sackler
Faculty of Exact Sciences, Tel Aviv University, Ramat Aviv, Tel
Aviv 69978, Israel and from Department of Physics, Ben Gurion
University, Beer Sheva 84105, Israel.

\medskip
\centerline{e-mail: aharony@post.tau.ac.il,
stauffer@thp.uni-koeln.de}

\medskip
Abstract: Yes and no. The size of the largest neighborhood in a
Barab\'asi-Albert scale-free network has strong fluctuations of
the order of the average value. The number of sites having exactly
ten neighbors increases linearly in the network size while its
relative fluctuations decrease towards zero if the number of sites
in the network increases from 1000 to ten million.
\bigskip

\begin{figure}[hbt]
\begin{center}
\includegraphics[angle=-90,scale=0.5]{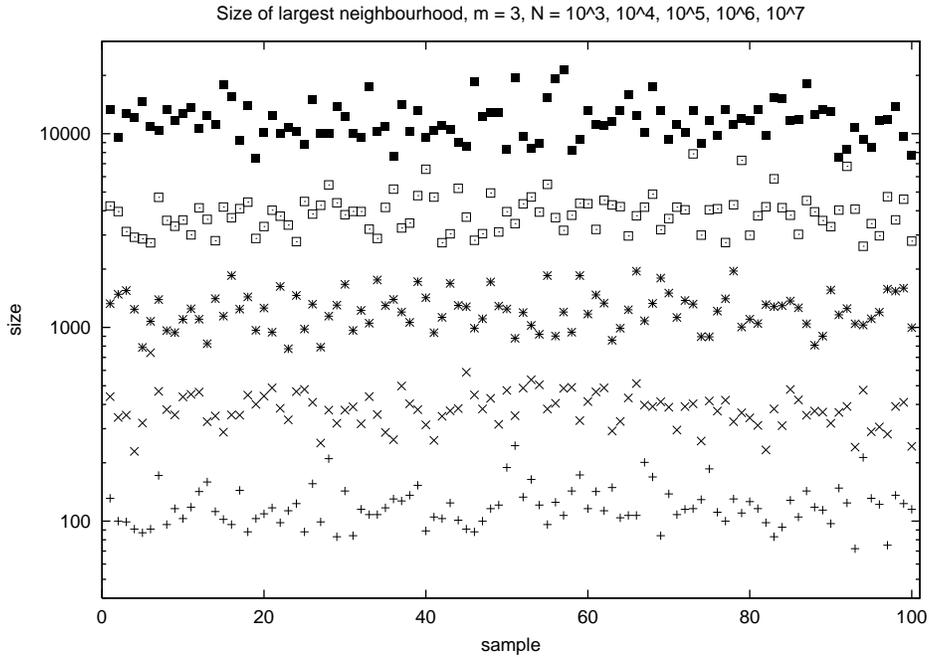}
\end{center}
\caption{ 100 examples for the size $k_1$ of the largest
neighborhood, with $10^3, \ 10^4, \, 10^5, \, 10^6, \, 10^7$ nodes
(bottom to top) added to the initial $m=3$ nodes. }
\end{figure}

\begin{figure}[hbt]
\begin{center}
\includegraphics[angle=-90,scale=0.4]{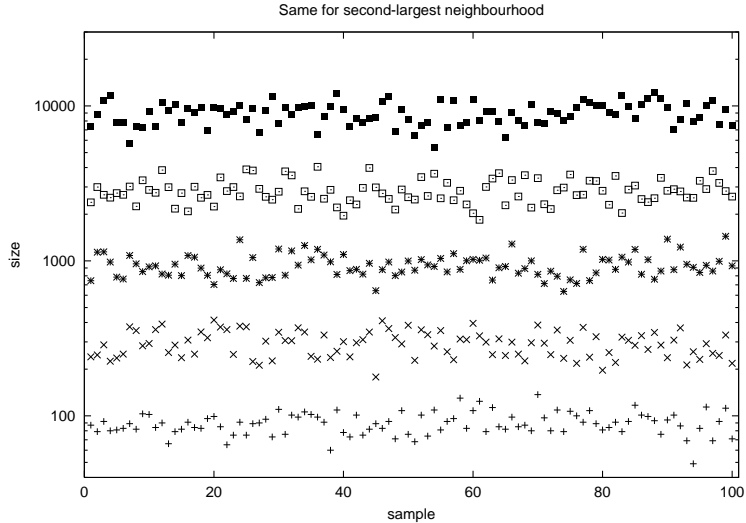}
\end{center}
\caption{ Same simulations as for Fig.1, now showing the size
$k_2$ of the second-largest neighborhood.}
\end{figure}

\begin{figure}[hbt]
\begin{center}
\includegraphics[angle=-90,scale=0.4]{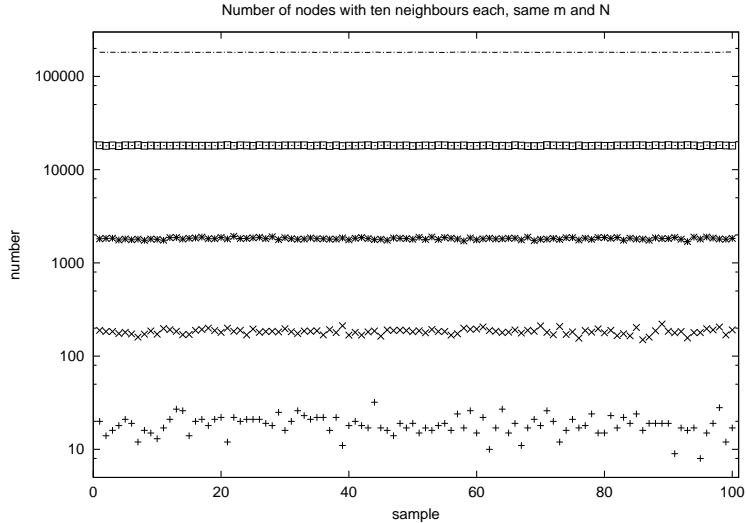}
\end{center}
\caption{ Same simulations as for Figs.1 and 2, now showing the
number of nodes having exactly 10 neighbors. }
\end{figure}

\begin{figure}[hbt]
\begin{center}
\includegraphics[angle=-90,scale=0.5]{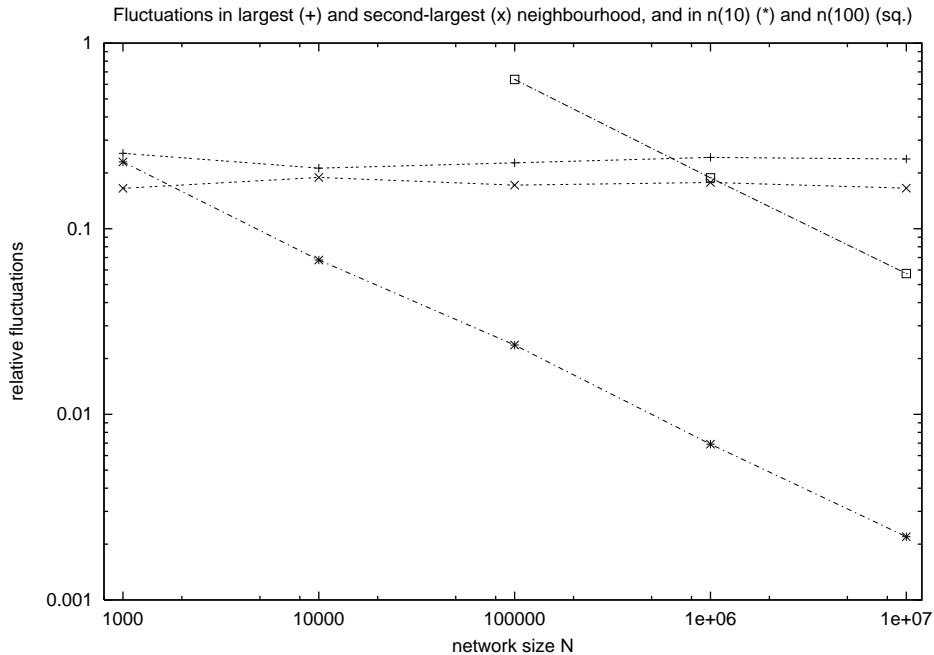}
\end{center}
\caption{ Same simulations as for Figs.1-3, now showing the
relative fluctuations of $k_1, \, k_2, \, n(10)$ and $n(100)$. }
\end{figure}

\begin{figure}[hbt]
\begin{center}
\includegraphics[angle=-90,scale=0.5]{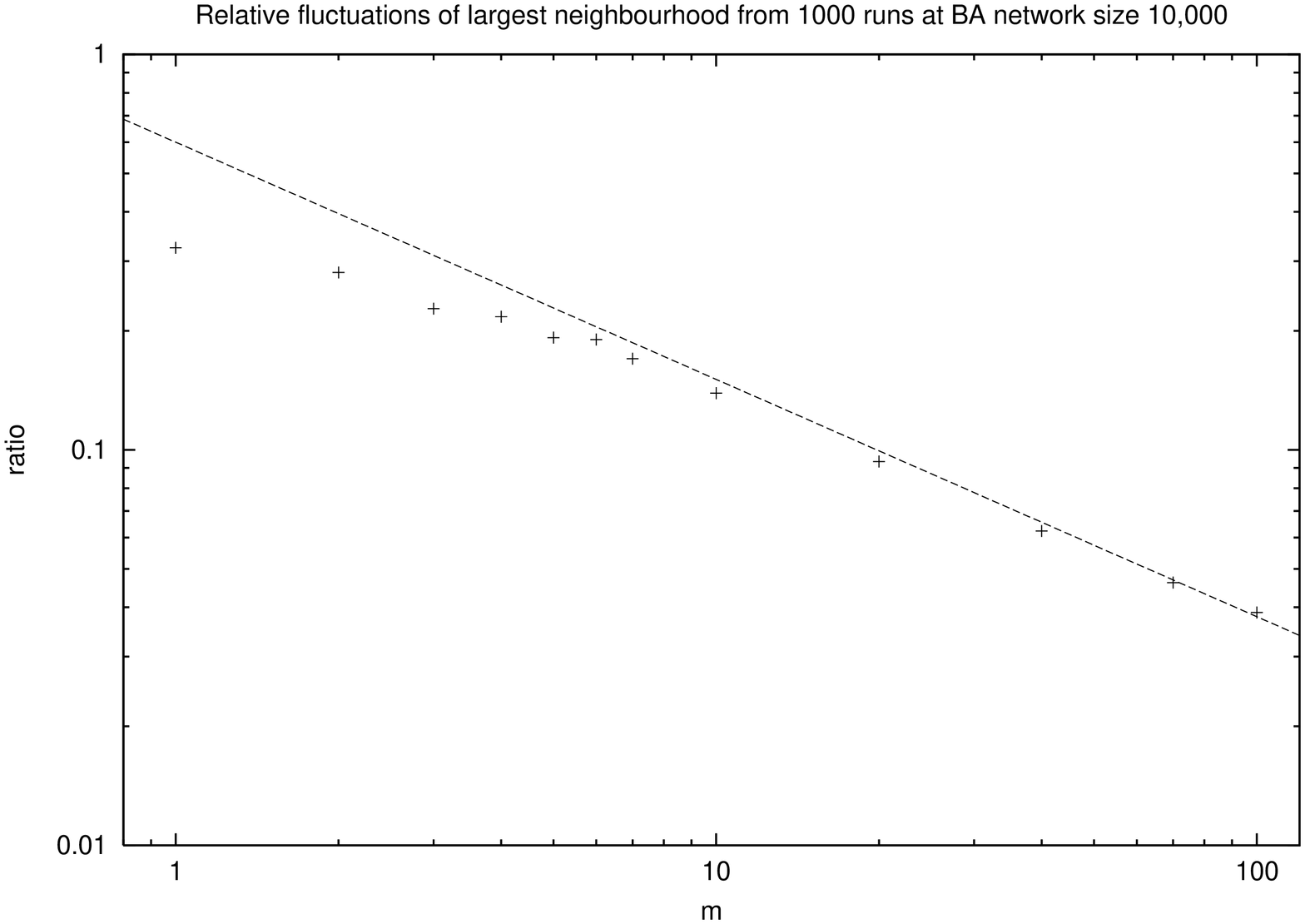}
\end{center}
\caption{ Log-log plot of relative fluctuations of $k_1$ versus $m$; 
the straight line has slope $-0.6$.}
\end{figure}

Self-averaging quantities have relative fluctuations which go to
zero for system size going to infinity, like the magnetization
away from the Curie point. Not self-averaging systems are found
e.g. at second-order phase transitions, like the magnetization at
the Curie point. Here we want to find out whether the scale-free
networks of Barab\'asi and Albert \cite{ba} are self-averaging.
The number of sites having a fixed number of neighbors is found to be strongly
self-averaging, confirming earlier results \cite{others}. However,
the  fluctuations in the two largest neighborhoods are seen to be 
proportional to these neighborhoods, implying large variations
among different realizations. Although the distribution of the
largest neighborhood was discussed in \cite{others}, the
consequences for self-averaging were not emphasized there.

To build these networks we start with $m$ sites all connected to
each other; we took $m=3$. Then $N$ more sites are added
one-by-one and each new site selects randomly exactly $m$ old
sites as neighbors, such that the probability for an old site to
be selected is proportional to the number of neighbors it has
already. Our network is undirected, i.e. if node A selected node B
as neighbor, then A is neighbor of B and B is neighbor of A. The
network is called scale-free since the number of sites (nodes)
having exactly $k$ neighbors decays as $1/k^3$. The size $k$ of
the largest neighborhood is called $k_1$, that of the
second-largest neighborhood is $k_2$, and the number of nodes
having exactly $k$ neighbors is $n(k)$.

We made 100 samples for $N = 10^3, \, 10^4, \, 10^5, \, 10^6$ and
show in our semilogarithmic figures without further analysis quite
clear results: The relative fluctuations of $k_1$ in Fig.1 and of
$k_2$ in Fig.2 remain strong even if we increase the network size
by four orders of magnitude. For $n(k=10)$, on the other hand, we
see in Fig.3 less and less relative fluctuations, as predicted in
\cite{doro}. Fig.4 confirms the visual impression from Figs.1-3 by
showing the relative fluctuations of the three discussed
quantities; the ones for $n(10)$ and $n(100)$ decay as $1/\sqrt
N$. The non-universal variation of the relative fluctuations for the
largest neighbourhood with varying $m$ (= number of neighbours 
selected by newly-added site) is shown in Fig.5.

We are not surprised about these results since they are similar to
cluster statistics in percolation (and presumably also in Ising)
models. The size of the largest cluster at the critical point
fluctuates strongly even in the thermodynamic limit \cite{book},
whereas the numbers of clusters of a fixed finite size get the
more accurate the larger the system is \cite{CSS}.

The German-Israeli Foundation supported this collaboration.

\end{document}